\voffset=.3in
\magnification=\magstep1
\baselineskip=1.5\baselineskip
\parskip=4pt plus 2pt minus 2pt 
%
\def\xx {|x\rangle\langle x'|}
\centerline {\bf A possible anisotropy in blackbody radiation viewed through} 
\centerline {\bf non-uniform gaseous matter}
\vskip 2pc
\centerline {T K Rai Dastidar\footnote *{Electronic address : 
mstkrd@mahendra.iacs.res.in}}
\vskip 1pc
\centerline {\it Indian Association for the Cultivation of Science, 
Calcutta 700032, India}
\vskip 3pc
\centerline {\bf Abstract}
\vskip 1pc
A non-local gauge symmetry of a complex scalar field, which can be
trivially extended to spinor fields,
was demonstrated in a recent paper (Mod. Phys. Lett. A{\bf 13}, 1265
(1998) ; hep-th/9902020). The corresponding covariant Lagrangian density 
yielded a new, non-local Quantum Electrodynamics.
In this Letter we show that as a consequence of this new QED, a
blackbody radiation viewed through gaseous matter
appears to show a slight deviation from the Planck formula, and
propose an experimental test to check this effect. We also show
that a non-uniformity in this gaseous matter distribution leads
to an (apparent) spatial anisotropy of the blackbody radiation. 
\vfill\eject
\centerline {\bf I. Introduction}
We have recently [1] shown how a non-local model of
Quantum Electrodynamics (QED), whose phenomenological form had been 
given earlier [2], results under a generalization of the usual
U(1) gauge transformation to a non-local one :
$$\exp(-ie\Lambda (x)) \to \exp(-ie\Lambda\xx).\eqno(1)$$
where $x\equiv$ ({\bf r}$,t$), $x'\equiv$ ({\bf r}$',t'$).
It was shown in [2] that a covariant derivative ${\cal D}_\mu=
\partial_\mu+d_\mu+ie{\cal A}_\mu\xx$ (where $d_\mu\equiv 
{\partial\over{\partial x'^\mu}}$) can be used to write down a covariant
Lagrangian for a complex scalar field $\phi(x)$ :
$${\cal L}=-{1\over 4}{\cal F}_{\mu\nu}{\cal F}^{\mu\nu}+
{\cal D}_\mu\phi^*{\cal D}^\mu\phi \eqno(2)$$
where
$${\cal F}^{\mu\nu}\equiv{\cal F}^{\mu\nu}\xx = (\partial^\mu+d^\mu)
{\cal A}^\nu\xx - (\partial^\nu+d^\nu){\cal A}^\mu\xx\eqno(3)$$
is the non-local electromagnetic field tensor. Under the gauge transformation
$${\cal A}_\mu\xx \to {\cal A}_\mu\xx+\partial_\mu\Lambda\xx
+d_\mu\Lambda\xx.\eqno(4)$$
${\cal F}^{\mu\nu}$ remains gauge invariant if and only if 
$$[\partial^\mu,d^\nu]=0.\eqno(5)$$
The relation (5) may therefore be considered
as the necessary and sufficient condition for the non-local gauge 
transformation (1) to be a symmetry of the field. This relation
(5), which implies that the two space-time points $x,x'$ are
completely independent (i.e. need not lie within the light cone of one
another), shows that the non-local correlation must be of the
Einstein-Podolsky-Rosen (EPR)-type. It was shown in [1] that :
\item {1.} A correlated two-photon absorption {\it linear in intensity} 
emerges as a consequence of this non-local symmetry of the radiation field. 
As mentioned in [1] (where references of related work in ordinary QED have
been given), experimental evidence of two-photon absorption of 
such kind has been obtained recently [3].
\item {2.} A non-thermodynamic Time's Arrow in the quantum (atomic) level
emerges as a necessary condition for energy conservation in such
correlated two-photon absorption processes, thus relegating the principle
of causality to a corollary of the principle of energy conservation.
\item {3.} The requirement for the EPR-type nonlocal correlation, which 
is in agreement with Bell's theorem, is also a necessary condition for 
energy conservation in interaction of matter with radiation involving 
a correlated two-photon exchange.

In this Letter it is shown that if we look, through
gaseous matter, at a blackbody radiation field (with an arbitrary
temperature T), then --- as a consequence of this correlated two-photon
absorption by the gas --- the spectrum of the radiation would appear to us 
to be slightly
deviated from the Planck formula, the amount of the deviation being
directly proportional to the density of the gaseous matter. Furthermore,
a non-uniform distribution of this matter would appear to us as an 
apparent spatial anisotropy of this radiation field temperature.
\vskip 2pc
\centerline {\bf II. Theory}
\vskip 1pc
The energy density of blackbody radiation at temperature T (deg K)
within a frequency band $\nu\to\nu + d\nu$ is given by the Planck
formula
$$E(\nu)d\nu=A\left ({u^3\over{e^u-1}}\right )d\nu,\quad {\rm where}\quad 
A={8\pi\over h^2}\left ({kT\over c}\right )^3,\quad u={h\nu\over kT}\cdot
\eqno(6)$$
The function $u^3/(e^u-1)$ is maximum at $u\simeq 2.821\thinspace 439$, i.e. 
the blackbody temperature T is related to the frequency $\nu_m$ at 
the peak by $T\simeq h\nu_m/(2.821\thinspace 439k)$.

We recall from [1] that when matter-radiation interaction satisfies
the condition of a second-order coherence, the symmetry of invariance
under a non-local gauge transformation becomes prominent and causes
two-photon absorption/s linear in the intensity. As a result,
when a blackbody radiation reaches an observer through a gaseous
absorber under circumstances (see discussion in section III)
that the above condition is fulfilled,
a coherent absorption of two photons of frequency ${1\over 2}\nu$ 
would be superposed as an absorption structure at a frequency $\nu$ 
upon the blackbody spectrum as viewed by the observer. 
Under the simplest assumption that the
absorption coefficient {\it z} is the same over the entire spectral range
of interest, the loss $F(\nu)$ as a consequence of the correlated 
absorption of two photons each of frequency ${1\over 2}\nu$ would be
given by
$$F(\nu)=zE(\nu/2)=zA\left ({y^3\over{e^y-1}}\right )\eqno(7)$$
where $y={1\over 2}u$. As a result the observed blackbody radiation 
density reduces to
$$G(\nu)=E(\nu)-F(\nu)=Af(u,z) \eqno(8)$$
where
$$f(u,z)={u^3\over{e^u-1}}-{{1\over 8}u^3\over{e^{{1\over 2}u}-1}}z\eqno(9)$$
\vskip 3pc
\centerline {\bf III. Results and discussion}
\vskip 1pc
Fig. 1 shows the functions $E(\nu),F(\nu)$ and $G(\nu)$ (times 
$1/A$) at a temperature T=2.728 deg K plotted against $\nu$, with an 
arbitrarily chosen absorption coefficient {\it z} = 5\%. It is evident 
that the peak of the $G(\nu)$-curve, which represents 
the intensity spectrum viewed by the observer through the gaseous matter,
is shifted slightly towards the left of the peak of the Planck curve 
$E(\nu)$, and this shift would appear to the
observer as {\it an apparent shift} $\Delta T$ {\it in the radiation 
temperature} --- in this particular
example the temperature shift turns out to be $\sim -0.039$ deg K, 
i.e. about 1.4\%. Sample calculations with different values of {\it z}
show that the temperature shift $\Delta T/T\sim 0.3z$ always.

It remains to ascertain exactly under what circumstances the correlated
two-photon absorption by the intervening gaseous matter can take place 
from the radiation field, so that the above effect can be observed. 
As shown in [1] (see \S 2 therein), 
the basic criterion is that the radiation field 
(either as generated, or as detected by the atomic electrons) should 
enjoy a large second-order coherence, i.e. it should be possible for 
two photons to be absorbed by the atomic electrons within a 
vanishingly small phase difference $\Delta\phi=\omega\delta t$, where
$\delta t$ is the time interval between the two photon-absorption events.
Essentially, this translates in practice to the condition that the
photon density be significantly higher than the electron density (the 
latter remaining, at the same time, high enough to make the photon 
absorption process viable).

A little reflection shows that this density criterion can be fulfilled under 
three circumstances: {\it first} --- in a laboratory setting --- when an atom 
is placed in a radiation field so 
intense that the photon density is substantially higher than the electron 
density ; {\it second} --- also a laboratory setting --- when an atom is 
placed in a phase-squeezed light
containing phase-correlated {\it signal} and {\it idler} photons ;
and {\it third}, in a cosmological setting where the overall photon density
is nine to ten orders of magnitude higher than the baryon or lepton density
and hence, in the millimetre-wave region as considered in the above example,
chances of two photon absorption events within vanishingly small
$\omega\delta t$ are high. For example, if two millimetre-wave photons
are absorbed within an interval of $\sim 0.1-1$ ps, then the phase gap
between them is $\ll 1$. The first and second laboratory settings
required for the second-order phase coherence and some of their experimental 
consequences have already been discussed in [1]~; the difference
between the laboratory settings and the cosmological setting is that,
whereas in the former the spatial distance between the two distinct
photon-absorption events is of the order of atomic/molecular dimensions,
in the latter the said distance can be of macroscopic scales.

Our exploratory calculations above have assumed that {\it z} is a constant.
Even without going into the detailed dynamics needed to calculate {\it z}
for any specific system, it is obvious that {\it z\/} depends on the
density of the absorbing gas. This allows for a very simple test for the
effect of an apparent shift in the observed blackbody radiation temperature
as predicted above. One simply needs a hollow cavity ``black body'' maintained 
at any arbitrary temperature, with a spectrophotometer mounted inside and 
filled with any desired gas, and record the spectrum at different pressures 
of the gas. So long as the pressure is sufficiently low so that the  
electron density is significantly lower than the photon density,  
the abovementioned second-order coherence condition applies, and the
{\it measured\/} blackbody temperature should show a variation with
the gas pressure as given in eqn.(8).
Furthermore, this pattern of linear variation of $\Delta T$ with 
{\it z} can be expected to hold only so long as the atomic/molecular 
density of the gaseous matter remains much less than the photon density 
of the radiation field.

Turning now to the cosmological setting,  if an observer looks at the 
cosmic microwave background radiation (CMBR) from
within a relatively dense gaseous atmosphere, then, as a consequence of
this correlated two-photon absorption (from the CMBR) by atmospheric
gases, he would find a 
relatively large deviation from the Planck curve, so long as the 
atomic/moleular density remains much less than the photon density ; 
however, viewing from {\it outside} the atmosphere, no such large 
deviation would appear
in his measurements. This seems to be corroborated by the difference 
between the balloon-borne measurements of the cosmic microwave background 
spectrum carried out by Woody and Richards [4] --- note in Fig.2 
therein the large deviation from the Planck curve at {\it twice} the peak
frequency --- and the measurements 
recorded by the {\it COBE\/} (Cosmic Background Explorer) satellite 
[5]. Of course, in the absence of any specific experimental data of the
type mentioned in the earlier paragraph, no precise reason for the
difference between the two sets of measurements can be given as yet, but
we can make the following speculations. The balloon measurements in [4]
were carried out at a float pressure of $\sim 2$ mbar, i.e. at a height of
about 120 km above sea-level, well up in the ionosphere. The peak in the CMBR 
spectrum in [4] is around 6 cm$^{-1}$, at a brightness of $\sim 10^{-11}$ 
W/cm$^2$, which corresponds to a photon flux $\sim 10^{11}$ cm$^{-2}$ 
sec$^{-1}$. (Dividing by $c$ gives a photon density $\sim 3$ cm$^{-3}$.)
{\it Twice of this photon energy\/} can correspond to any of the following :

\item{i)} spacing between high-lying  Rydberg levels (such as $ns^2-n'p^2,
np^2-n'd^2\dots$) of
Helium [6], where $n$ and $n'$ are of order $\sim 40$ and can be equal ;
\item{ii)} spacing between high-lying levels of He$^+$ ;
\item{iii)} spacing between rotational sublevels of high vibrational
levels of electronically excited states of H$_2$ or H$_2^+$.

Assuming an electron temperature of $\sim 1.5\times 10^4$ deg K [7] and a 
Helium concentration of around 5 ppm in dry air, use of the Saha formula [8] 
for excited-state population of He and He$^+$ gases at the stated
pressure in local thermal equilibrium yields a density of order $\sim 
10^{-4}-10^{-5}$ cm$^{-3}$ ; this is 
several orders of magnitude lower than the CMBR photon density. It is
difficult to estimate the corresponding figure/s for the Hydrogen molecules,
in view of the uncertainty in the concentration of water vapour in the
ionosphere. Nevertheless, it seems safe to assume that the population density
of the rovibrationally excited levels would be much smaller than the 
relevant photon density. We can therefore say that the energy levels of
excited Helium and hydrogen and/or their ions as present in the upper
atmosphere (at $\ge 100$ km altitude) {\it allow a coherent two-photon 
absorption from the CMBR, giving rise to an 
apparent shift in the spectrum\/}
as given in Fig.\thinspace 1~; experimental investigations as spelt out 
earlier in the paper would be required to make a quantitative comparison
with the data in [4] possible.
(Note that we do not expect any noticeable effect in
{\it ground-based observations\/}, because of the lack of ionized and/or
excited species.)

Furthermore, if the density of any gaseous matter intervening between
the observer and the radiation be non-uniform in space, i.e. different 
along different directions with respect to the observer, then this 
apparent shift in the blackbody radiation tempterature would also vary 
along different directions --- in other words, a non-uniformity in the 
absorbing matter distribution would lead to an {\it apparent spatial 
anisotropy} in the observed temperature. It is to be noted that this
effect is independent of the Doppler anisotropy caused by the observer's
motion through the radiation field (e.g. [9]). Of course, our
theory does not cover neutrinos which are supposed to make up the {\it
dark matter\/} of the universe, and hence cannot be used to fully account
for the spatial anisotropies in the CMBR~; work is in progress to 
include the electroweak interactions within our model.

In another forthcoming work, we intend to show how, in principle, it is 
possible for the cosmic background radiation to impart the atomic-level
time's arrow (as derived in [1] and mentioned towards the beginning of 
this paper) to all particles in the universe that interact with the
electromagnetic field.
\vskip 2pc
The author is indebted to Dr Krishna Rai Dastidar for fruitful 
discussions and for critically reading the manuscript.
\vfill\eject
\centerline {\bf REFERENCES}
\vskip 1pc
\item{1.} T K Rai Dastidar and Krishna Rai Dastidar, Mod. Phys. Letts. 
A{\bf 13}, 1265 (1998) ; {\it Errata} A{\bf 13}, 2247 (1998). Also 
available in the internet as~: xxx.lanl.gov/abs/hep-th/ 9902020.
\item{2.} T K Rai Dastidar and Krishna Rai Dastidar, Abstracts of XI
Int. Conf. At. Phys. (Paris, 1988), p. I-23 ; Krishna Rai Dastidar {\it in}
``Advances in Atomic and Molecular Physics'', ed. M S Z Chaghtai (Today's
and Tomorrow's Publishers, New Delhi, 1992) p. 49.
\item{3.} N P Georgiades, E S Polzik, K Edamatsu, H J Kimble and
A S Parkins, Phys. Rev. Lett. {\bf 75}, 3426 (1995)
\item{4.} D P Woody and P L Richards, Phys. Rev. Lett. {\bf 42}, 925
(1979)
\item{5.} J C Mather et al, Astrophys. J. {\bf 420}, 439 (1994)
\item{6.} T K Rai Dastidar, unpublished calculations.
\item{7.} See e.g. {\it Astrophysical Formulae\/} by K R Lang (Springer, 
1978), Table 9.
\item{8.} M N Saha, Proc.\thinspace Roy.\thinspace Soc.\thinspace 
Lond. {\bf A99}, 135 (1921)
\item{9.} P J E Peebles and D T Wilkinson, Phys. Rev. {\bf 174}, 2168
(1968) 
\vfill\eject
\centerline {\bf Figure caption}
\vskip 1pc
\item {Fig.\thinspace 1.} The Planck function $E(\nu)$ (full line), 
coherent two-photon absorption loss $F(\nu)$\break (dashed line) with an 
absorption coefficient {\it z} = 5\%, and the resultant
intensity function $G(\nu)$ (dotted line) plotted against the 
frequency $\nu$. (T=2.728 deg K.) All the functions have been 
scaled down by the factor {\it A} (eq.(6)) 
to yield dimensionless quantities.

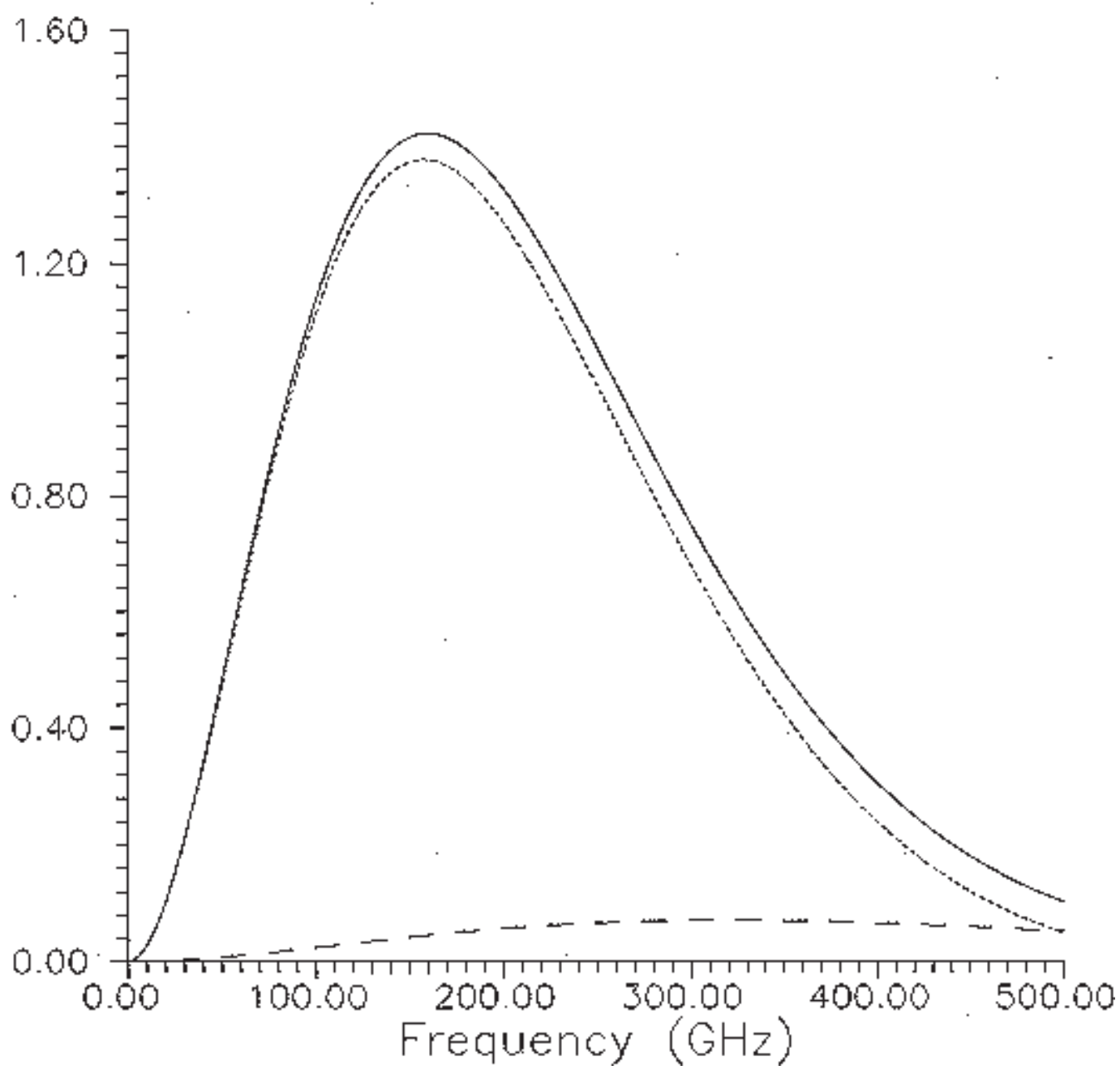
\bye